\documentclass{nature}

\usepackage{graphicx}
\makeatletter
\let\saved@includegraphics\includegraphics
\AtBeginDocument{\let\includegraphics\saved@includegraphics}
\renewenvironment*{figure}{\@float{figure}}{\end@float}
\makeatother
\usepackage{bm}% bold math
\usepackage[version=3]{mhchem}
\usepackage{natbib}
\usepackage{epsfig}
\usepackage{amsmath}
\usepackage{bbold}
\usepackage{float}

\usepackage{xcolor}

\makeatletter
\newcommand*{\rom}[1]{\expandafter\@slowromancap\romannumeral #1@}
\setcitestyle{super}

\title{Decoherence in Molecular Electron Spin Qubits:  Insights from Quantum Many-Body Simulations}

\author{Jia Chen$^{1,2,3}$, Cong Hu$^4$, John F. Stanton$^{2,3,5}$, Stephen Hill$^{3,6}$, Hai-Ping Cheng$^{1,2,3}$, \& Xiao-Guang Zhang$^{1,2,3}$}

\begin{document}

\maketitle
\begin{affiliations}
\item Department of Physics, University of Florida, Gainesville, FL 32611, USA
\item Quantum Theory Project,  University of Florida, Gainesville, FL 32611, USA
\item Center for Molecular Magnetic Quantum Materials
\item Department of Physics, University of Connecticut, Storrs, CT 06269, USA
\item Department of Chemistry, University of Florida, Gainesville, FL 32611, USA
\item Department of Physics and National High Magnetic Field Laboratory, Florida State University, Tallahassee, Florida 32306, USA

\end{affiliations}
\begin{abstract}
Quantum states are described by wave functions whose phases cannot be directly measured, but which play a vital role in quantum effects such as interference and entanglement. The loss of the relative phase information, termed decoherence, arises from the interactions between a quantum system and its environment. Decoherence is perhaps the biggest obstacle on the path to reliable quantum computing. Here we show that decoherence occurs even in an isolated molecule although not all phase information is lost via a theoretical study of a central electron spin qubit interacting with nearby nuclear spins in prototypical magnetic molecules. The residual coherence, which is molecule-dependent, provides a microscopic rationalization for the nuclear spin diffusion barrier proposed to explain experiments. The contribution of nearby molecules to the decoherence has a non-trivial dependence on separation, peaking at intermediate distances. Molecules that are far away only affect the long-time behavior. Because the residual coherence is simple to calculate and correlates well with the coherence time, it can be used as a descriptor for coherence in magnetic molecules. This work will help establish design principles for enhancing coherence in molecular spin qubits and serve to motivate further theoretical work. 
\end{abstract}

Electron and nuclear spins provide a natural realization of qubits and have been identified as potential building blocks for quantum technologies.\cite{Cory1997, Gershenfeld1997} Dynamical considerations are such that electron spins hold a clear advantage in terms of operational speed,\cite{Warren1997} but suffer in terms of decoherence due to stronger interactions with their environment. Electron spins can be accessed through magnetic materials. Magnetic molecules are uniquely attractive because synthetic methodologies allow optimization of their quantum properties by tuning both the immediate coordination environment of the electronic qubit, i.e., the underlying spin-Hamiltonian, and the peripheral molecular structures that mediate interactions with the environment.\cite{Gaita-Arino2019, Atzori2019}

Spin dynamics can be characterized in terms of longitudinal and transverse relaxation times: T$_1$ and T$_2$ respectively. For molecular qubits at low temperatures, T$_2$ typically limits the number of quantum operations that can be performed before phase coherence is lost. A successful strategy to extend T$_2$ involves modifying the environment through dilution of the magnetic molecules in diamagnetic solvents.\cite{Gaita-Arino2019} Examples of molecules with very long T$_2$ times that were realized in this way include \ce{[Cu(mnt)2]^{2–}}(Ref.\citen{Bader2014}) and \ce{[V(C8S8)]^{2-}} (Ref.\citen{Zadrozny2015}). In these dilute cases at low temperatures, experimental studies agree that the main source of electron spin decoherence involves relatively weak couplings to nuclear spins.\cite{Lenz2017}

\begin{figure}
\includegraphics[width=\linewidth,clip]{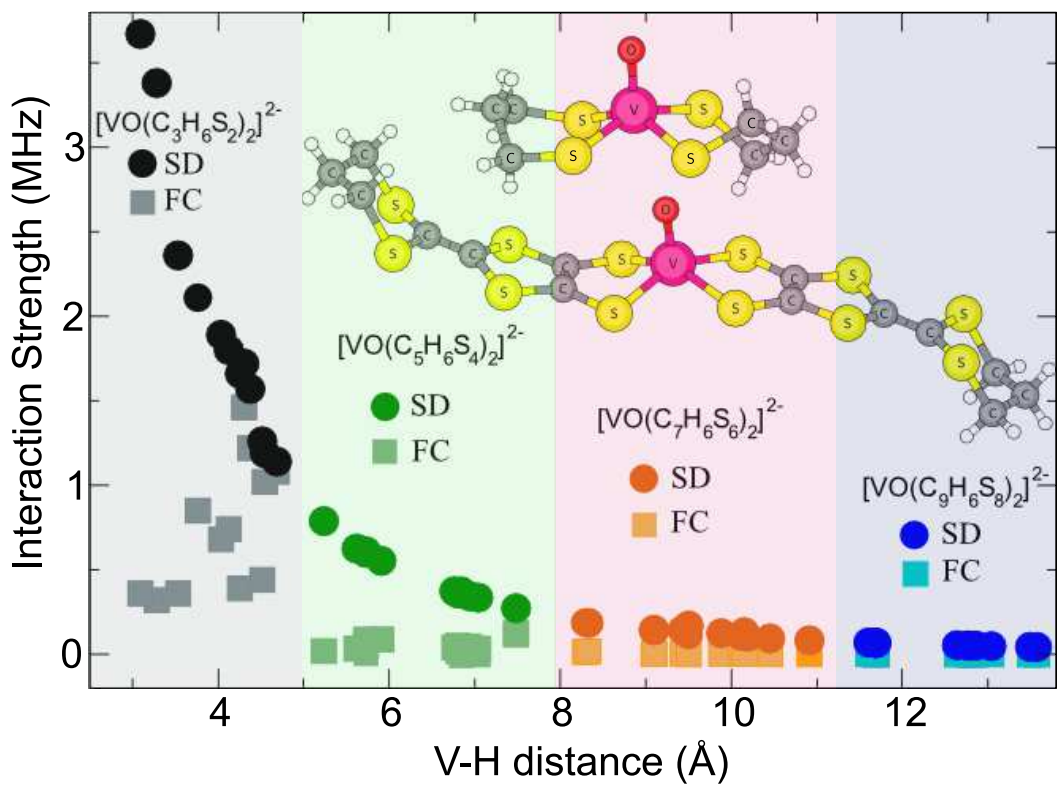}
\caption{ Electron-proton hyperfine coupling as a function of V-$^1$H separation for the four vanadyl complexes, including the isotropic Fermi-contact (FC) interaction from B3LYP calculations and the anisotropic through-space spin-dipolar (SD) contribution. The molecular structures of \ce{[VO(C3H6S2)2]^{2-}} and \ce{[VO(C7H6S6)2]^{2-}} are shown above the data: V - pink; O - red; S - yellow; C - gray; H - white.} \label{H_hyperfine}
\end{figure}

The relationship between molecular structure and T$_2$ was first illuminated, \cite{Graham2017} and later expanded upon\cite{Graham2017_2} by Freedman {\it et al}. In the earlier work, a series of four vanadyl complexes, \ce{(Ph4P)2[VO(C3H6S2)2]}, \ce{(Ph4P)2[VO(C5H6S4)2]}, \ce{(Ph4P)2[VO(C7H6S6)2]}, and \ce{(Ph4P)2[VO(C9H6S8)2]} (see Fig.~\ref{H_hyperfine}), with increasing distance between the magnetic vanadium ion and the spin-active hydrogen nuclei, were synthesized and T$_2$ was measured in dilute frozen solutions by monitoring the decay of the Hahn echo. The study found that smaller complexes with stronger electron-nuclear interactions actually had longer T$_2$ times. This counterintuitive result was rationalized on the basis of a nuclear spin diffusion barrier.\cite{Graham2017} 

One of the first proposed spin decoherence mechanisms is nuclear spin diffusion mediated by energy conserving spin flip-flop processes.\cite{Bloembergen1949} However, this process is suppressed for nuclear spins that are close to electron spins, because their effective Zeeman energies are significantly detuned from their neighbors in comparison to less proximate spin pairs. This region within which nuclei have limited contribution to decoherence is described by the diffusion barrier.\cite{Ramanathan2008} Experiments with different settings have reported the sizes for barriers ranging from 3 to 17 \AA.\cite{Schmugge1965,  Wolfe1973, wolfe1973_2, Goldman1965, Ramakrishna1966, Tse1968, Tan2019}  Because of the connection between decoherence and structure,\cite{Graham2017} a first-principles theory for spin decoherence in molecular qubits that can be connected to molecular structure is clearly a desirable research objective. Such a theory, with the promise to gain insights and to make predictions about spin decoherence, would also provide practical benefits to the experimental community.

For all four vanadyl complexes studied in this work, each molecule contains twelve distal hydrogen atoms, believed to be the main source of electron spin decoherence at low temperatures. We calculated hyperfine coupling tensors between electron spin and nuclear spins of both $^{51}$V and $^1$H using density functional theory.
\cite{Neese2003} Comparison of calculations and experimental measurements of coupling to $^{51}$V can be found in \textbf{Methods} section; reasonable agreement can be achieved. Therefore, we lean on calculations for hyperfine coupling of $^1$H, and the results are summarized in Fig.~\ref{H_hyperfine}. Paramagnetic spin-orbit (PSO) terms are smaller than $0.05~$MHz in all cases, and are neglected in this work. Spin-dipolar (SD) interactions decay as $ 1 / d^3$, where $d$ is the distance between the central vanadium ion and any given hydrogen atom. Meanwhile, the Fermi-contact (FC) term is insignificant for all but the smallest complex. This is to be expected since the electron spin density is mostly confined to the vanadium {\it d} orbitals. Therefore, as the V-$^1$H distance increases, the direct electron-proton contact rapidly diminishes. Consequently, the spin-dipole interaction dominates the results, as seen in Fig.~\ref{H_hyperfine}, and this can be computed on the basis of the atomic coordinates.

Decoherence can be understood in terms of the dynamics of a closed quantum system that includes the central spin (the qubit) and a finite number of  environmental, or bath spins.\cite{Yao2006,Yang2008} Correlations between the central spin and the bath spins result in leakage of quantum information to the environment.\cite{Zurek2003} We have adopted a bottom-up approach to study this decoherence, induced by electron-nuclear hyperfine coupling in molecular qubits. Starting from a single molecule, we extend the approach to include the effects of more distant molecules with active nuclear spins surrounding the central vanadyl complex. Since the dipolar interaction is a tensorial property, the spin will evolve uniquely for different molecular orientations, i.e., the decoherence process is intrinsically anisotropic. Therefore, in order to reproduce the results of experiments performed on ensembles of randomly oriented molecules in frozen solutions, we employed an algorithm\cite{Kuffner2004} to sample a uniform distribution of molecular orientations. Also, initial nuclear wave functions were generated by uniformly distributed random coefficients. Results reported here are averaged over both random orientations and initial wave functions.

\begin{figure}
\includegraphics[width=\linewidth,clip]{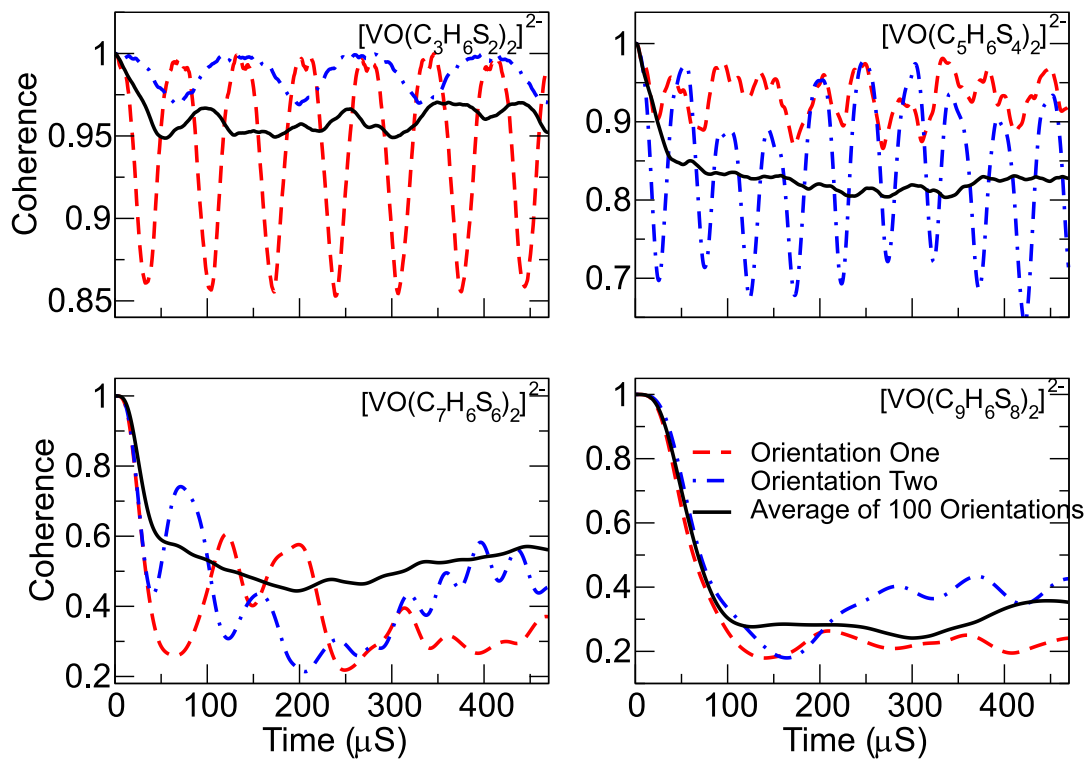}
\caption{ Decay of Hahn echo for a single molecule calculated using the pair-correlation approximation, averaging over 20 random initial wave-functions and 100 random orientations. The external magnetic field is set to $1.0$ T.} \label{orientation}
\end{figure}

In this work, the dynamics of nuclear spins are simulated by solving the time-dependent Schr\"{o}dinger equation, and their effects on electron spin decoherence are accounted for via the cluster-correlation expansion scheme,\cite{Yang2008} which divides decoherence into contributions from clusters of nuclear spins. The lowest order pair-correlation approximation,\cite{Yao2006} in which each cluster consists of two nuclear spins, is already able to capture the nuclear spin flip-flop process. Inclusion of larger clusters will lead to more accurate results by including correlations of more bath spins. The time dependence of electron spin coherence, which is quantified by the magnitudes of the off-diagonal elements of the reduced density matrix, is determined after forward and backward time evolution,\cite{Yang2008} in a manner that is consistent with Hahn echo measurements of T$_2$.\cite{Graham2017}  Results of simulations using the pair-correlation approximation\cite{Yao2006} are shown in Fig.~\ref{orientation}. 

For the smaller complexes, \ce{[VO(C3H6S2)2]^{2-}} and \ce{[VO(C5H6S2)2]^{2-}}, simulations for a single molecular orientation reveal oscillating behavior, with very different oscillation amplitudes and periods for different orientations. Upon averaging over 100 uniformly distributed orientations, we find that, after an initial decay, the coherence plateaus at a nonzero ``residual coherence'' value. This can be understood as arising from dephasing of oscillations corresponding to different molecular orientations after the initial sharp decay, which essentially represents the ensemble average coherence over multiple periods of oscillations. For the larger complexes, \ce{[VO(C7H6S6)2]^{2-}} and \ce{[VO(C9H6S8)2]^{2-}}, a clear decay is seen for just a single molecular orientation, and the residual coherence decreases with increasing molecular size. The reduced coherence and the more complex oscillations for the individual orientations indicate enhanced participation of protons in the decoherence process as their separation from the central spin increases.   

\begin{figure}
\includegraphics[width=\linewidth,clip]{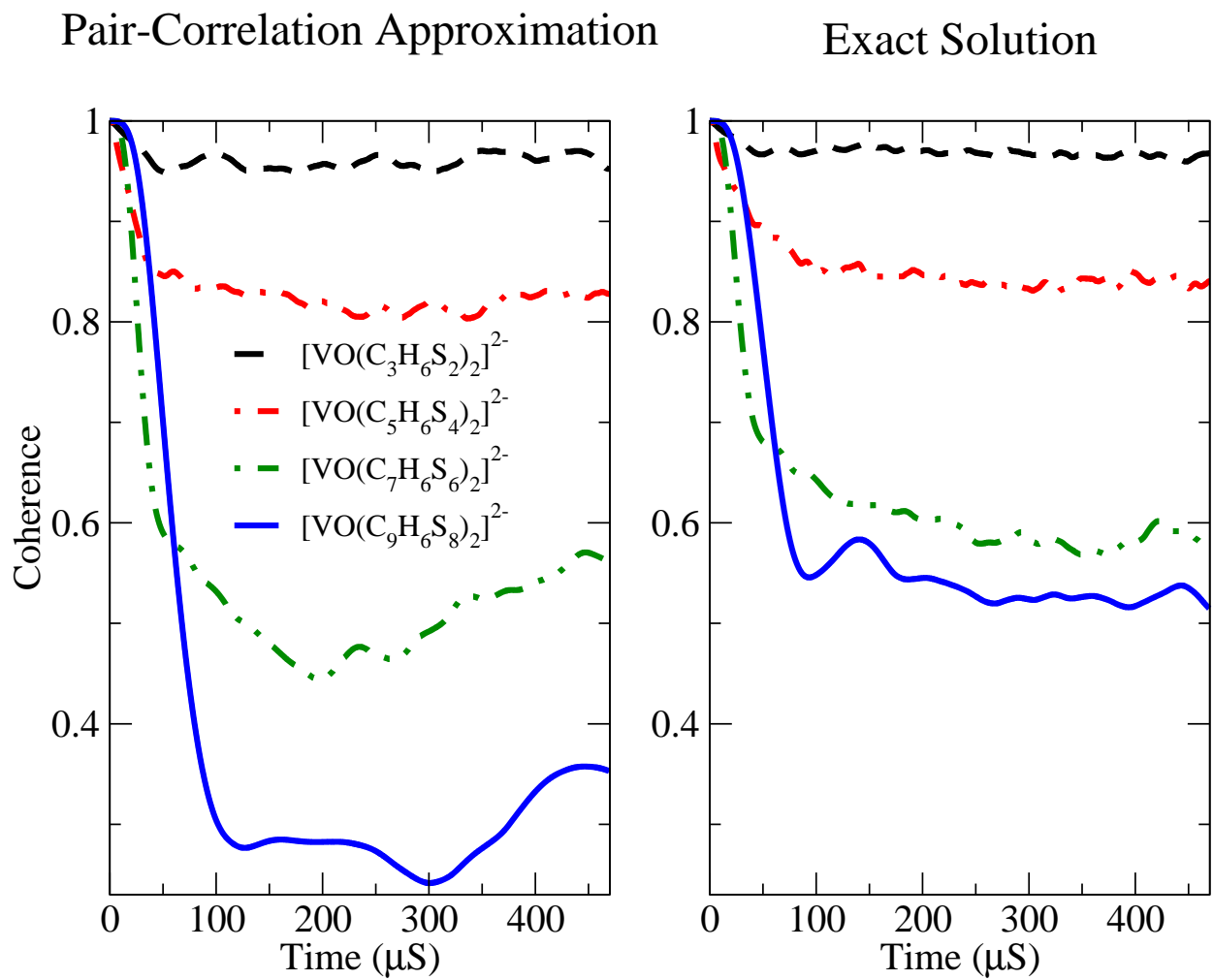}
\caption{ Single molecule Hahn echo decay calculated for the four complexes by averaging over 20 random initial wave-functions and 100 random orientations, based on the pair-correlation approximation (left panel) and exact solution (right panel).} \label{correlation}
\end{figure}

A direct comparison of the four complexes under study can be seen in Fig.~\ref{correlation}. As noted above, it is clear that the decoherence is incomplete for a single molecule with just twelve protons, and that the level of residual coherence decreases as the size of molecule increases. This suggests a quantum information ``leakage bottleneck'', dictated by the finite number of protons and their freedom to participate in the decoherence process. In essence, protons in close proximity to the central spin have a limited capacity to decohere the central spin. This rationalises the observed trend of T$_2$ seen in experiments,\cite{Graham2017} with larger residual coherence tracking longer T$_2$ values.

To study multi-spin correlation effects on decoherence in these vanadyl complexes, we need to go beyond pair-correlation approximation. The quantum dynamics of all twelve $^1$H nuclear spins in one molecule can be solved exactly via matrix diagonalization. In the cluster-correlation expansion scheme,\cite{Yang2008} the exact solution for one molecule corresponds to one cluster with twelve spins. Such exact solutions can be found in the right panel of Fig.~\ref{correlation}. Important qualitative features of the decoherence observed from the pair-correlation approximation calculations are reproduced in the exact solutions: ``residual coherence'' corresponds to the plateau reached after initial decay, and the level of residual coherence decreases as the size of the molecule increases. Therefore, we will use the pair-correlation approximation to gain additional insights for larger systems. However, the residual coherence from the exact solutions is somewhat larger, especially for larger molecules, suggesting that correlations involving multiple spins must be considered in order to achieve quantitative agreement.

Since single-molecule residual coherence can be calculated from the molecular electronic structure alone using the procedure described here, and magnetic dilution can isolate molecular qubits in solutions, residual coherence can be used as a theoretical proxy for T$_2$ in this setting. Identifying such descriptors for surface chemical reactions and in the search for new catalysts has been a driving force for recent developments in the field of theoretical surface chemistry and catalysis.\cite{Norskov2011}  It is enticing to propose that residual coherence can play a similar role for molecular spin qubits. Moreover, residual coherence provides a microscopic picture for the nuclear spin diffusion barrier. After adopting diffusion barrier of 6.6\AA~ to 4.0\AA, as reported in experiments,\cite{Graham2017} we can connect residual coherence to diffusion barrier, and the value of residual coherence turns out to 0.9 in this case. Residual coherence can be viewed as a calculable measure of how much certain nuclear spins contribute to decoherence. Since it is a continuous variable, residual coherence contains more information than diffusion barrier as a single cutoff.

The effects of environmental decoherence were further studied by considering hyperfine coupling to nuclear spins outside of the molecule on which the electron spin resides. In the experiments, the vanadyl complexes were diluted in the deuterated solvents d$_7$-dimethylformamide/d$_8$-toluene,\cite{Graham2017} in which d$_7$ and d$_8$ signify 7 and 8 protons have been replaced by deuterons in each molecule. Thus, only the vanadyl complexes and the counter ion, \ce{Ph4P+}, contain protons which are the most destructive nuclei for electron spin coherence. Since the negatively charged vanadyl and positively charged \ce{Ph4P+} counter ion experience an attractive Coulomb interaction, dilution may not be able to separate them effectively. Consequently, \ce{Ph4P+} provides an additional source of environmental protons that can couple to the electronic spin, and we performed further simulations to elucidate their effect. 

\begin{figure}
\includegraphics[width=\linewidth,clip]{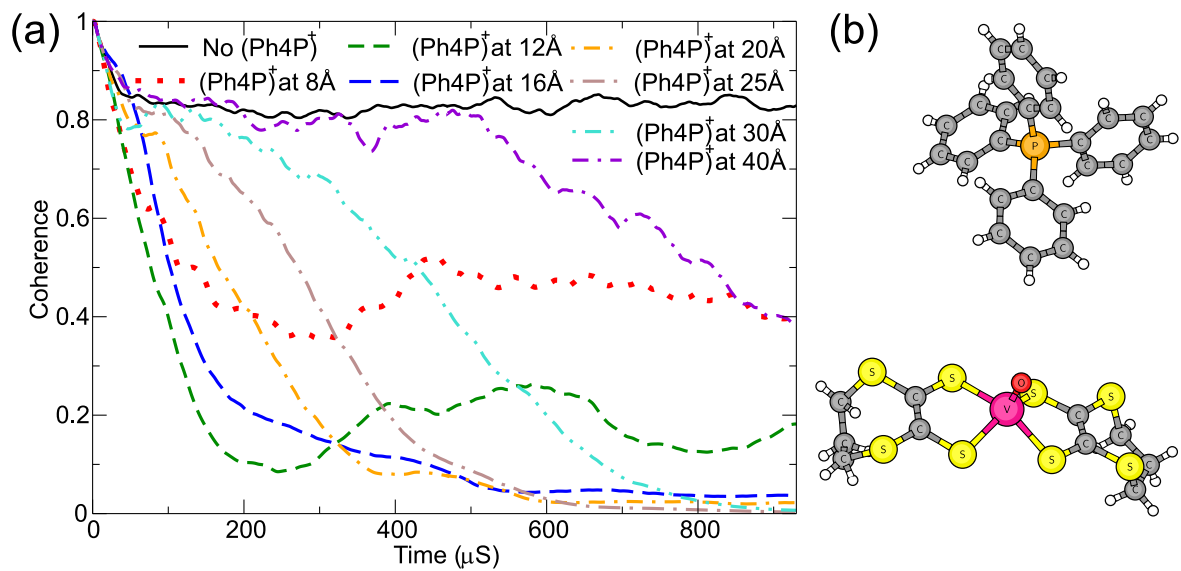}
\caption{(a) Electron Hahn echo decay for \ce{[VO(C5H6S4)2]^{2-}} with one \ce{Ph4P+} counter ion after averaging over 100 molecular orientations and 20 initial wave-functions at various distances. (b) Molecular model for \ce{[VO(C5H6S4)2]^{2-}} and \ce{Ph4P+} at a separation of 10\AA: V - pink; O - red; S - yellow; C - gray; H - white; P - orange.} \label{pph4}
\end{figure}

The simplest model has one vanadyl complex and one \ce{Ph4P+} counter ion. We then studied the coherence as a function of the distance between \ce{Ph4P+} and the vanadyl. Results for \ce{[VO(C5H6S4)2]^{2-}} are shown in Fig.~\ref{pph4}, and are representative of what was found for all four complexes. In panel (a), we see that the counter ion facilitates decoherence tremendously, with a single \ce{Ph4P+} at 8\AA ~ away reducing the residual coherence from 0.8 to 0.4. As we move \ce{Ph4P+} ion from 8\AA~ to 12\AA~ away from vanadyl complex, the residual coherence decreases even more. This indicates that the effectiveness of environmental molecules for decoherence has a non-linear dependence on their distance to central spin. Within a short distance, being closer actually compromises environmental molecules' ability to decohere. As we move \ce{Ph4P+} farther away, decoherence shows the behavior of a two-step process. This becomes apparent when \ce{Ph4P+} is farther than 30\AA~ away. After the initial decay, a recurrence is seen, and the second and much slower decoherence process commences. Initial decay is similar to the behavior of a single molecule, because inflection point can be found at the value of the residual coherence without environmental molecules. Therefore, we can conclude that the initial decay is due to nuclear spins within \ce{[VO(C5H6S4)2]^{2-}}: when \ce{Ph4P+} is farther than 30\AA,~ it makes no contribution to the short-time decoherence.

Observations from the simulations provide perspective regarding more effective dilution schemes to extend T$_2$. Since counter-ions are an important source of decoherence, it may be possible to tune or even optimize the T$_2$, if their proximity to the molecular qubit can be modulated by the choice of solvent. Realization of such tunability will require knowledge of the frozen solution structure. Simple options, such as using deuterated or aprotic polar solvents, can obviously still impact T$_2$.

To summarize, electronic structure calculations and quantum many-body simulations offer valuable insights into electron spin decoherence in vanadyl molecular qubits. Electron spins in all four complexes are localized on the \ce{V^{4+}} ion. Hyperfine interaction tensors were calculated as analytic derivatives using density functional theory. Electron spin decoherence from the dominant spin-dipole interaction mechanism was studied via quantum many-body simulations based on the cluster-correlation expansion. For a single vanadyl molecule, decoherence is always incomplete; it reaches a plateau of residual coherence whose magnitude scales inversely with the distance between the electron spin and the distal $^1$H nuclei. Further effects of the environment on the decoherence were studied by including an additional molecular counter-ion in the simulations containing nuclear spin. The residual coherence shows a pronounced dependence on the intermolecular distances. Molecules far away generally do not contribute to short-time decoherence. Agreement between the calculated residual coherence and the trend observed experimentally suggests that the residual coherence is an appropriate descriptor for T$_2$ in molecular spin qubits. This is convenient, as it is a simple function of molecular electronic structure and is amenable to straightforward computational characterization. Thus, it may be that a predictive model of spin decoherence is now at hand, which clearly has profound potential for the study of molecular qubits.  

\begin{methods}

\section{First principle calculations of 
hyperfine coupling tensors}

Hyperfine coupling involves the interaction between electron and nuclear spins. The corresponding interaction Hamiltonian can be written as:
\begin{align}
H_K^{hf} = \mathbf{S} \cdot \mathbf{A_{K}} \cdot \mathbf{I_{K}} , 
\label{hyperfine_tensor}
\end{align}
where $\mathbf{S}$ is the electron spin; $\mathbf{I_K}$ is the nuclear spin of nucleus K and $\mathbf{A_K}$ is the associated hyperfine coupling tensor. Like other interaction parameters in EPR and NMR effective Hamiltonians, the hyperfine coupling tensor can be calculated based on first principles electronic structure theories via analytical derivative techniques.\cite{Trygve1999} Equation \ref{hyperfine_tensor} suggests that the hyperfine coupling tensor can be viewed as the derivative of total energy with respect to the electron and nuclear spins. It consists of three parts,\cite{Helgaker1999} 
\begin{align}
\mathbf{A_K^{tot}} = \mathbf{A_K^{SD}} + \mathbf{A_K^{FC}} + \mathbf{A_K^{PSO}} ,
\label{three_parts}
\end{align}
where $\mathbf{A_K^{SD}}$ represents the spin-dipole interaction, which is the classical interaction between two magnetic dipoles, and $\mathbf{A_K^{FC}}$ represents the Fermi contact interaction due to the overlap between the unpaired {\it S} electron wave-function and the nucleus. The Fermi contact interaction is isotropic, thus $\mathbf{A_K^{FC}}$ is diagonal. The third term, $\mathbf{A_K^{PSO}}$, is a paramagnetic spin-orbit term, a relativistic effect that is usually much smaller than the spin-dipole and the Fermi contact terms.

Density functional theory (DFT) was employed to study the electronic structure, particularly hyperfine couplings, of those vanadyl complexes which are responsible for the decoherence of electron spins. Specifically, we used the quantum chemistry software ORCA\cite{Neese2012, Neese2018} to calculate hyperfine coupling tensors.\cite{Neese2003} Commonly-used exchange-correlation functionals, BP\cite{Becke1988, Lee1988}, at generalized-gradient approximation (GGA) level, and B3LYP\cite{Becke1993, Vosko1980, Stephens1994} at the hybrid functional level were deployed. In order to obtain accurate Fermi contact interactions, a core polarized basis set [CP(PPP)]\cite{Neese2002} was used for Vanadium and a triple-$\zeta$ basis set TZVP\cite{Schafer1994} was used for other elements. 

To test the calculations of hyperfine coupling for this series of complexes, we first compare results with available experimental measurements done on $^{51}$V. The calculated hyperfine coupling tensors between electron spin and $^{51}$V are listed in Table.~\ref{calc_hyper}. In the principal axis system, the largest component of the spin-dipole interaction is slightly larger than the Fermi-contact interaction; the paramagnetic spin-orbit term is more than an order of magnitude smaller than both the spin-dipole and Fermi-contact interaction. For the four complexes, hyperfine coupling tensors are very similar, with variations among molecules less than 10 percent. Due to the localization of the electron spin around the vanadium ions in all four complexes, hyperfine coupling is not strongly influenced by atoms outside the first coordination shell of V$^{4+}$ ion. Since more localized electron spin tends to produce stronger hyperfine coupling with central ions, hybrid functional B3LYP gives larger coupling tensors than generalized-gradient approximation level functional BP.

Comparisons between calculation and EPR measurement can be found in Table.~\ref{calc_exp}. It shows that, B3LYP underestimates the hyperfine coupling slightly; for the largest component, the discrepancy is between 10\% to 20\%. Since BP predicts even smaller interactions, hybrid functional B3LYP gives better hyperfine coupling tensors for the series of complexes compared to EPR measurements. In general, DFT calculations give reasonable hyperfine couplings of $^{51}$V. Because we do not have experimental data for $^1$H, we are relying on the values given by B3LYP calculations for coupling tensors of $^1$H.

\begin{table}
\caption{ Components of hyperfine coupling tensors (in MHz) for $^{51}$V (S=$\frac{7}{2}$) of vanadyl complexes in principal coordinates }\label{calc_hyper}
\begin{tabular}{lcccccccc}
\hline
Complex & Method &A$^{SD}_{11}$ & A$^{SD}_{22}$ & A$^{SD}_{33}$ & A$^{FC}_{ISO}$ & A$^{PSO}_{11}$ & A$^{PSO}_{22}$ & A$^{PSO}_{33}$ \\
 \hline
\ce{[VO(C3H6S2)2]^{2-}} & BP & 90.1 &  80.1 & -170.2 & -140.3 &  -6.1 & -6.4 & -12.8  \\
  & B3LYP & 100.3 & 90.3 &  -190.6 & -162.7 & -7.7 &              -8.0 & -16.8  \\
\ce{[VO(C5H6S4)2]^{2-}} & BP & 82.7 & 81.6 & -164.3 &  -134.4 & -5.2 & -5.5 & -13.1  \\
  & B3LYP &  92.9 & 91.7 & -184.7 & -159.6 & -6.6 &              -6.9 &  -17.2  \\
\ce{[VO(C7H6S6)2]^{2-}} & BP & 89.7 & 74.4 & -164.2 & -134.5 & -5.3 & -6.0 & -14.7  \\
  & B3LYP &  100.2 &  85.8 & -186.1 & -161.0 & -6.8 & -7.5  & -18.8  \\
\ce{[VO(C9H6S8)2]^{2-}} & BP & 87.2 & 75.2 & -162.4 & -132.9 & -4.9 & -5.5 & -13.1  \\
  & B3LYP & 96.3 & 86.0 & -182.3 & -157.4 & -6.2 & -6.8 & -16.6 \\
 \hline
\end{tabular}
\end{table}

\begin{table}
\caption{ Total hyperfine coupling tensors (in MHz) for $^{51}$V of vanadyl complexes in principal coordinates from B3LYP calculation and EPR measurement.\cite{Graham2017} Only absolute value of calculated numbers are reported, assuming experiments can not tell sign of measured numbers }\label{calc_exp}
\begin{tabular}{lccccc}
\hline
Complex & A$^{exp}_{\perp}$ & A$^{exp}_{\|}$ & (A$^{calc}_{11}$+ A$^{calc}_{22}$)/2 & A$^{calc}_{33}$ \\
 \hline
\ce{[VO(C3H6S2)2]^{2-}} & 125 & 418 & 75 & 370 \\
\ce{[VO(C5H6S4)2]^{2-}} & 120 & 395 & 74 & 361 \\
\ce{[VO(C7H6S6)2]^{2-}} & 129 & 416 & 75 & 366 \\
\ce{[VO(C9H6S8)2]^{2-}} & 129 & 417 & 73 & 356 \\
 \hline
\end{tabular}
\end{table}

\section{ Quantum Many-Body Dynamics }
In this work, the electron spin is the system of interest (the central qubit) and hydrogen nuclear spins represent the environment. We aim to study the quantum dynamics without invoking additional approximations necessary for stochastic theory or master equations. Such a theory has been developed and successfully applied to quantum dots\cite{Yao2006} and other nanoscale systems, as reviewed in Ref.~\citen{Liu2007}. Following the same approach, we evolve the electron spin starting from its initial state, which is assumed as a superposition state of $\alpha |\uparrow\rangle + \beta | \downarrow \rangle$. The initial wave-function for all the nuclear spins is denoted by $|J(0)\rangle$. The nuclear spins evolve with time under a Hamiltonian that depends on the state of the electron spin. We denote the nuclear spin Hamiltonian by $\hat{H}^+$ and $\hat{H}^-$, corresponding to the electron spin states $|\uparrow\rangle$ and $|\downarrow\rangle$, respectively. At a later time $t$, the nuclear spin wave-function is $|J^{\pm}(t)\rangle = e^{-i\hat{H}^{\pm} t / \hbar} |J(0)\rangle$. Then the coherence can be defined as the off-diagonal element of the electron spin reduced density matrix,
\begin{align}
L(t) = \frac{\rho_{\uparrow\downarrow}(t)}{\rho_{\uparrow\downarrow}(0)} = \langle J^-(t) | J^{+}(t) \rangle = \langle J(0) | e^{i\hat{H}^-t/\hbar} e^{-i\hat{H}^+t/\hbar}| J(0) \rangle .
\label{coherence}
\end{align}
This allows decoherence to be studied as a function of the time evolution of the nuclear spin wave-functions.

Hamiltonians corresponding to the electron spin up and down for nuclear spins can be decomposed into three terms, a Zeeman term $-\gamma_n \hat{\mathbf{S}} \cdot \mathbf{B}$, where $\gamma_n$ is the gyromagnetic ratio for $^1$H, an intrinsic spin-spin interaction $\hat{H}_{in}$, and extrinsic interaction $\pm \hat{H}_{ex}$,
whose sign is determined by the electron spin,
\begin{align}
\hat{H}^{\pm} = -\gamma_n \hat{\mathbf{S}} \cdot \mathbf{B} + \hat{H}_{in} \pm \hat{H}_{ex}.
\label{Hpm}
\end{align}
The intrinsic nuclear spin-spin interaction is the magnetic dipole-dipole interaction. We used secular approximation\cite{Levitt2001} for this type of interaction,
\begin{align}
\hat{H}_{in} = \sum_{i\neq j} -\frac{\mu_0}{4\pi} \frac{\gamma_n \gamma_n \hbar}{r^3_{ij}} \frac{3 \cos^2\theta_{ij}-1}{2}(2\hat{S}_i^z\hat{S}_j^z-\hat{S}_i^x\hat{S}_j^x-\hat{S}_i^y\hat{S}_j^y),
\label{secular}
\end{align}
where $\theta_{ij}$ is the angle between vector connecting spin $i, j$ and external magnetic field $\mathbf{B}$. This term can account for the spin flip-flop process since $\hat{S}_i^x\hat{S}_j^x+\hat{S}_i^y\hat{S}_j^y = \frac{1}{2}(\hat{S}_i^+\hat{S}_j^-+\hat{S}_i^-\hat{S}_j^+)$.

The extrinsic interaction consists of two parts,
\begin{align}
\hat{H}_{ex} = \sum_i a_i S^z_i + \sum_{i\neq j} \frac{a_i a_j}{2B} (S_i^+ S_j^- + S_i^- S_j^+).
\label{ex}
\end{align}
The first term is the Ising part of the hyperfine interaction; the second term is the off-diagonal interaction between nuclear spins mediated by the hyperfine, which has a similar origin as the RKKY interaction.\cite{Ruderman1954} Since the electron spin is localized on vanadium ions in these complexes, the parameter $a_i$ can be calculated based on the distance between hydrogen and vanadium and the angle between the vector connecting hydrogen and vanadium and the magnetic field,
\begin{align}
a_i = \frac{\mu_0}{4\pi} \frac{\gamma_n \gamma_e \hbar}{r^3_{i}} (3 \cos^2\theta_{i}-1) .
\label{ai}
\end{align}

The nuclear spin wave functions can be evaluated exactly for a small number of spins. Approximations are needed when we have a large environment. The pair-correlation approximation\cite{Yao2006}, which treats different spin-pairs as uncorrelated, is generally good for large systems. For small systems like a molecule, correlated motion involving more than two spins may contribute to decoherence,\cite{Yang2008} and more careful treatment may be necessary. One successful extension to include those effects is the cluster-correlation expansion method.\cite{Yang2008} In this method, contributions from larger clusters can be systematically added. The limitation is the number of spins in each cluster that we can solve exactly. Moreover, arbitrary controlling pulses can be implemented in this method by the contour-time-dependent Hamiltonian.\cite{Yang2008} Here we report calculations of Hahn echo, since this is the approach used to measure T$_2$ for this series of complexes in experiments.\cite{Graham2017}    

\section{Data availability}
The data that support the findings of this study are available from the corresponding author upon reasonable request

\end{methods}

\begin{addendum}
 \item This work was supported as part of the Center for Molecular Magnetic Quantum Materials
(M2QM), an Energy Frontier Research Center funded by the US Department of Energy, Office of
Science, Basic Energy Sciences under Award DE-SC0019330. C.H. thanks the support from the
NSF REU program under grant DMR-1852138. Work performed at the National High Magnetic Field Laboratory is supported by the NSF (DMR-1644779) and the State of Florida
 \item[Competing Interests] The authors declare that they have no
competing financial interests.
 \item[Correspondence] Correspondence and requests for materials should be addressed to X-G. Z. (email: xgz@ufl.edu).
\end{addendum}

\bibliography{Collection}

\end{document}